\documentclass[doublecol]{epl2}
\usepackage{epsfig} 

\title{Determination of tunneling charge via current measurements}
\author{Ines Safi\inst{1} and Eugene V. Sukhorukov\inst{2}}
\institute{
\inst{1} Laboratoire de Physique des Solides, Universit\'e
Paris-Sud, 91405 Orsay, France\\
\inst{2} D\'epartement de Physique Th\'eorique, Universit\'e de
Gen\`eve, CH-1211 Gen\`eve 4, Switzerland}


\abstract{We consider a tunnel junction between two arbitrary non-linear systems in any dimension, which can be  different. We show that the tunneling charge can be detected using three alternative methods based on current measurements. {Besides being technically easier compared to noise measurements,
these methods present valuable advantages: they do not require the knowledge of the underlying models, and some
are accessible in the experimentally convenient  low-voltage regime, where heating effects are reduced}. The first method is based on the AC conductance, while the two others are based on photo-assisted current (PAC) and can be implemented for any time-dependence of the tunneling amplitude. These are promising for edge states in the regime of the fractional quantum Hall effect (FQHE): the Hamiltonian does not have to be specified and can incorporate non-universal interactions between the edges, and it is more convenient to use an AC gate voltage rather than an AC bias. These methods apply for instance to weak barriers in 1-D systems, Superconductor-Insulator-Normal (SIN) or graphene-like structures.}

\begin{document}

\maketitle

One of the most exciting hallmarks of electronic interactions are elementary excitations with non-integer charges or with a double charge in the super-conducting state \cite{khlus,sanquer_noise}. Searching for methods to measure them has attracted tremendous efforts on both the theoretical and experimental level. Fractional charges have even been predicted recently in graphene-like structures \cite{frac_graphene} or topological insulators \cite{frac_topo}. A rich variety has been mainly offered by the FQHE, where quasi-particles are predicted also to obey fractional statistics. Their measurement plays a crucial role as it may shed light on the underlying physics. While for simple filling factors (defined as the dimensionless ratio between the number of electrons
and that of the flux quanta) $\nu=1/(1+2n)$, with $n$ integer, the charge is simply $q=\nu e$, it was claimed that for the Jain series \cite{jain_series}, $\nu=p/(1+2np)$, the fundamental charge is $e/(1+2np)$. Recently, there has been a resurgence of interest in the non-abelian $\nu=5/2$ fractional state, whose different descriptions have a common prediction: the existence of a charge $e/4$.
{Its detection, being one of the crucial steps toward revealing non-abelian statistics \cite{non_abelian_theo}, has not been conclusive yet
\cite{heiblum_one_fourth,radu_exp,heiblum_bunch}.} Indeed, for any filling factor, theory and experiment have not been fully reconciled so far.

Theoretical suggestions to determine the charge of carriers has been mostly based on the Poissonian
formula for the noise $S$ at sufficiently weak current $I$,
\begin{equation}\label{noise}
S=qI,
\end{equation}
obtained for some specific systems \cite{kane_fisher_noise}, and used to access the charge
of the carriers experimentally \cite{charge_hall}. It turns out that this formula does not depend on
{the choice}
of the underlying model \cite{S_I,ines_eugene}. However, its experimental applications encounter a number of difficulties. First, noise is more difficult to measure compared to current. Second, Eq.\ (\ref{noise}) requires the condition of high voltage compared to temperature, $qV\gg T$ (we set $k_B=1$ and $\hbar=1$) \cite{note_condition}, which can cause heating effects in experiments. In Heiblum's group \cite{heiblum_two_third}, where the regime of higher voltages is investigated, this equation has been replaced by a conjectured one: that for non-interacting particles with charge $q$. This has led to results which have not been fully understood, even for simple fractions. There has also been a theoretical proposal to use the photo-assisted noise to detect the charge at simple fractions \cite{crepieux_photo}, a result which we can indeed generalize to any filling factors and arbitrary systems in any dimension \cite{ines_eugene}. But still such a proposal is based on noise and requires $qV\gg T$.

 On the other side, there are some experiments based on conductance measurements \cite{radu_exp}. Their disadvantage is that they rely on a particular model for the edges, thus on factors such as the edge reconstruction and Coulomb interactions. For instance, the exact structure of edge modes in ``hole conjugate'' sates, at fillings $2/3$, $3/5$, etc., remains still an unsolved issue \cite{kane_fisher_jain}.

Here we propose alternatives to measure the charge which avoid one or, simultaneously, many of these limitations: they are based on current measurements, and some of them are possible in the low voltage regime, {which reduces} heating. The important aspect of these results is their universality,
i.e., the independence of the model as long as the system is kept
in a weak tunneling regime. We exclude only the treatment of a super-current,
{which in an Superconductor-Insulator-Superconductor (SIS) structure}
can be suppressed by a magnetic field. More generally, we can allow for two different systems, for an STM and another system to probe, and for coupling to any type of electromagnetic environment. Our computation applies to both tunneling and weak backscattering in correlated 1-D systems.
The charge $q$ enters the equations (\ref{reG}--\ref{secondIbis})
 in many different
ways, which allows us to propose a variety of methods to detect
it. Some of them require zero-bias anomaly (ZBA), such as in 1-D systems and the FQHE, systems with disorder and interactions and/or coupled to an ohmic environment.

Two of our proposals are based on the Photo-Assisted Current (PAC) \cite{tunnel,photo_theo,photo_exp,crepieux_photo}. Indeed, we make another important achievement: it gives a generalization of the theory of Tien-Gordon\cite{tunnel} intended to deal with Photo-Assisted Tunneling (PAT) in SIS junctions, and
that by Tucker \cite{tucker_two}. We extend these theories to treat:
\vspace*{-0.3cm}
\begin{list}{\labelitemi}{\leftmargin=1em}\addtolength{\itemsep}{-0.5\baselineskip}
\item
A general Hamiltonian required only to conserve the charges in both systems (in the absence of tunneling), thus allowing for Coulomb interactions between them. This is of great importance in the FQHE.
\item
A tunneling charge different from $e$, without specifying the explicit form of the tunneling term.
 \item Differentials of the current with respect to the Fourier components of an ``effective'' voltage {kept} finite.
 \item Arbitrary periodic dependence on time of the voltage or/and the tunneling amplitudes.
  \end{list}
The latter is  easier to settle in the FQHE. Indeed, time-dependent barriers in one dimension have been studied previously \cite{LL_time}, where the current was expressed explicitly, depending on the underlying model, {thus have a different spirit and aim.}
Besides, we offer a unified background which contains all those different results, and make the bridge with the PAT.
\begin{figure}[htbp]
\vspace{0.3cm}
\begin{center}
\includegraphics[width=7cm]{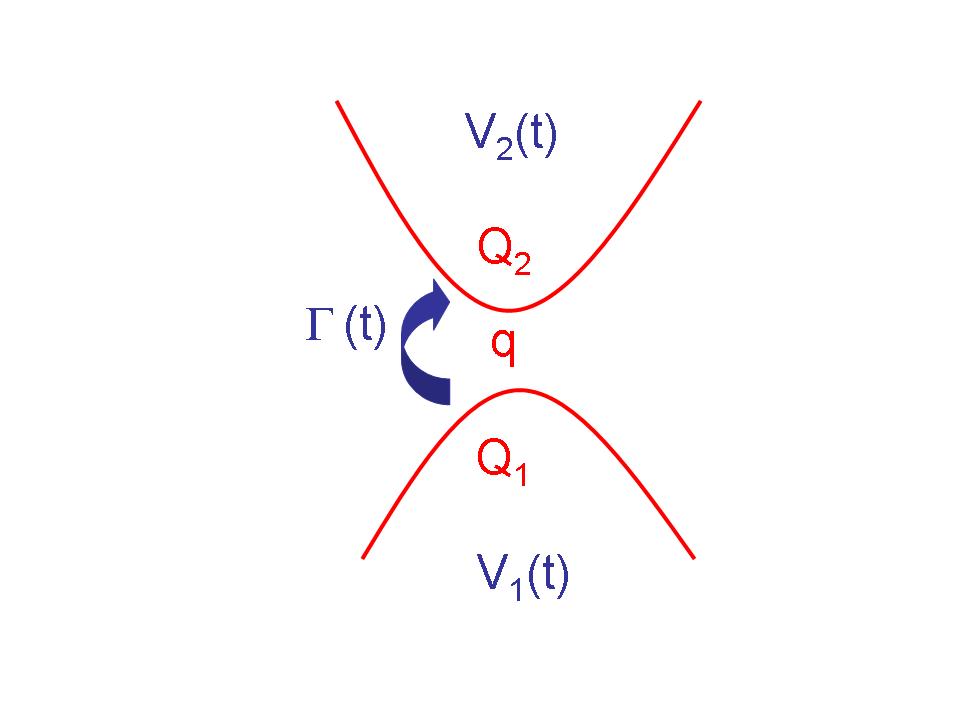}
\vspace{0.15in} \caption{\small Tunnel junction between two systems 1 and 2 in any dimension with arbitrary interactions or disorder,
 subject to time-dependent potential $V_{1,2}(t)$. They can be either similar or different, such as fractional edge states with equal or different filling factors, or SIN structures. One can also view one system as a probe, such as STM tip or a Fermi liquid coupled to edge states or to a quantum dot. Mutual Coulomb interactions are allowed, we require only $Q_1$ and $Q_2$ to commute with the total Hamiltonian $\mathcal{H}_0$ in the absence of tunneling.}
\label{setup}
\end{center}
\end{figure}

{\bf Model:}
We start from two (identical or different) systems whose Hamiltonian $\mathcal{H}_0$ incorporates interactions or
 disorder which do not need to be specified.
 In particular, we do not require $\mathcal{H}_{0}$ to be decoupled into two Hamiltonians, thus can include mutual Coulomb interactions.
 The only requirement for $\mathcal{H}_0$ is that it does not change the total
charges $Q_{1}$ and $Q_{2}$: (see Fig.\ \ref{setup})
$[Q_l,\mathcal{H}_{0}]=0$ for $l=1,2$. We describe a gauge invariant coupling to
the applied potentials $V_1(t)$ and $V_2(t)$ by $\mathcal{H}_Q=Q_1V_1(t)+Q_2V_2(t)$,
 and tunneling between the two systems by $\mathcal{H}_T$. Thus the total Hamiltonian reads :
\begin{equation}
\mathcal{ H}=\mathcal{H}_0+\mathcal{H}_Q+\mathcal{H}_T.
\label{Hamiltonian}
\end{equation}

 We allow for an arbitrary dependence of the tunneling amplitudes $\Gamma_{k_1k_2}{(t)}$ on the states $k_1$ and $k_2$ on both sides, as well as for their time-dependence, assuming it only to be identical for the different states, i.e., the dimensionless function: $\gamma (t)=\Gamma_{k_1,k_2}(t)/\Gamma_{k_1,k_2}(0)$ does not depend on $k_1,k_2$. This is reasonable if time dependence were to be controlled by a nearby gate. Therefore, denoting by $\mathcal{T}$ the tunneling operator at time $0$, we have:
\begin{equation}
\mathcal{H_T}=\gamma(t)\mathcal{T}+\gamma^*(t)\mathcal{T}^{\dag},
\label{tunneling}
\end{equation}
where we do not have to write the explicit form of $\mathcal{T}$. For the perturbative computation carried out later, we require only that $\left<\mathcal{T}(t)\mathcal{T}(0)\right>=0$ in the absence of tunneling \cite{heisenberg}, which implies that the super-current is zero, or has to be negligible compared to that of quasi-particles. In an SIN structure, one has $q=2e$
{(respectively, $q=e$) at low (respectively, high) energies compared to the gap}. {In the FQHE as well, $q$ can depend on energy scales.}

The operator $\mathcal{T}$ transfers a charge $q$, thus translates the charge
difference {$Q=(Q_2-Q_1)/2$ by $q$.
Therefore} it contains, without any loss of generality, an exponential of a displacement operator conjugate to $Q$: $\mathcal{T}=\mathcal{T'}e^{iq\Phi}$, where $[\Phi,Q]=i$, and $\mathcal{T}'$ commutes with both $\Phi$ and $Q$. This is in particular suited to include coupling to an electromagnetic environment or even to another quantum system. Using the scalar phase $\phi$ defined as
$\dot\phi=V_1(t)-V_2(t)\equiv V(t)$, and the standard time-dependent unitary transformation
$\mathcal{U}(t)=e^{-i Q\phi(t)}$, one has $
\mathcal{U}(t)\mathcal{T}\mathcal{U}^{\dagger}(t)=e^{iq\phi(t)}\mathcal{T}$.
This allows us to absorb $\mathcal{H}_Q$ by applying
$\mathcal{H}\to U\mathcal{H}\mathcal{U}^{\dag}-iU\partial_t\mathcal{U}^{\dag}=\mathcal{H}_0+\mathcal{H}_T(t)$,
so that only $\mathcal{H}_T$,
being gauge non-invariant, acquires additional time-dependence:
\begin{eqnarray}
\mathcal{H}_T(t)&= &E(t)e^{iq Vt}\mathcal{T}+E^*(t)e^{-iqVt}\mathcal{T}^{\dagger}.
\label{tunneling-t0}\\E(t)&=&e^{iq\tilde{\phi}(t)}|\gamma(t)|\label{E(t)}.
\end{eqnarray}
The phase $\tilde{\phi}(t)$ and $V$ emanate from both $V(t)$ and $\arg\gamma(t)$. More precisely,
the acquired phase $\arg [\gamma(t)]/q+\phi(t)=Vt+\tilde{\phi}(t)$ is decomposed so that its time
derivative, yielding an ``effective'' voltage, contains
a DC part $V$ and a time-dependent one $\tilde{V}(t)=\partial_t{\tilde\phi}(t)$.

Now it is only $\mathcal{H}_T(t)$ which drives the system
from an equilibrium state. Treating it as a perturbation, we apply the standard procedure of expanding the evolution operator in the interaction representation, keeping only the
first non-vanishing contribution to $I$, {the average of the tunneling current operator:
\begin{equation}\label{IT}
\hat{I}(t)=iE(t)e^{iq Vt}\mathcal{T}-iE^*(t)e^{-iqVt}\mathcal{T}^{\dagger}.
\end{equation}}
 The non-equilibrium dynamics is contained in the factor $E(t)E^*(t')$,
while
\begin{equation}\label{Xpm}
 X_{-}(t)=\langle\left[\mathcal{T}^{\dagger}(t),\mathcal{T}(0)\right]\rangle
 \end{equation}
  becomes invariant under time translation: now the average
  {$\langle \ldots\rangle\equiv{\rm Tr}\rho_0(\ldots)$}
  is taken with respect to the unperturbed equilibrium density matrix
  {$\rho_0\propto e^{-\beta \mathcal{H}_0}$}
  (all our results are valid at arbitrary $T=1/\beta$). We denote by $X^R(t)=\theta(t)X_-(t)$, such that $Re X^R(\omega)=X_-(\omega)$.
Let us first consider the case when $\tilde{V}(t)=0$ and $|\gamma(t)|=1$, thus $E(\omega)=2\pi\delta(\omega)$.
Then, formally, we have: $I(qV)=-{ q}X_-(-qV)/{\pi}.$ Notice that $I$ is measured as a function of $V$, but we keep $q$ as $V$ appears always multiplied by $q$, when compared to frequency [see Eq.\ (\ref{reG}) for example]. It is easy to show, by a spectral decomposition, that $I(0)=0$, which conforms with our assumption that we ignore the super-current. However, $I$ is not necessarily an odd function of $V$.
We can as well show that the zero-frequency noise is given by Eq.\ (\ref{noise}) universally.
 {The finite-frequency (FF) noise will be expressed in Ref.\ \cite{ines_eugene}}.

{ {\bf Out-of-equilibrium AC conductance:}} $G(qV;\omega)$ \cite{AC} can be expressed either through an exact formula \cite{gavish,safi_ac} adapted for tunneling \cite{ines_eugene}, or from a perturbative computation:
$2\pi\omega G(qV;\omega)= q^2\left[X^R(-qV+\omega)-X^{R*}(-qV-\omega)-2i {\rm Im}\ X^R(-qV)\right].$ An interesting consequence is that its dissipative part
\begin{equation}\label{reG}
{\rm Re}\,G(qV;\omega)=\frac {q}{2\omega}[I(qV+\omega)-I(qV-\omega)]\,,
\end{equation}
can be expressed solely in terms of the DC current. This is obviously satisfied in the
Ohmic regime, but becomes less trivial in the nonlinear regime.
{Similar expression, restricted by $q=e$,
has been derived in \cite{tucker_two,worsham_ac}.}
Note, {the expressions (\ref{noise}) and (\ref{reG})} are the sort of the fluctuation-dissipation relations, being derived perturbatively with respect to tunneling. Indeed, ${\rm Re}\, G$ yields the asymmetric part of the FF noise \cite{gavish,safi_ac,ines_eugene}, whose full expression {can be expressed also in terms of the DC current} \cite{tucker_two,ines_eugene}. Recall that Eq.(\ref{reG}) is valid at arbitrary $T,V,\omega$. We will now propose three regimes where it can be used to measure $q$.

The form (\ref{reG}) offers a first alternative: $q$ can be determined through the prefactor and simultaneously by the structure of the r.h.s.\ at $\omega=\pm qV$. For instance, if $I$ shows up a ZBA, ${\rm Re}\,G(qV;\omega)$ gets similar anomalies translated at $V=\pm \omega/q$. This method requires that one of the translated voltages $V\pm \omega/q$ is in the domain of non-linearities of $I$, and to reach frequencies $\omega\simeq V$.

Experimentally, low $\omega$ are easier to access, which yields the second regime, $\omega\ll qV$, for which Eq.\ (\ref{reG}) reduces to:
\begin{equation}
\label{expansion_G}{\rm Re}\,G(qV;\omega)- G(qV)\simeq \frac{\omega^2}{6q^2}\frac{d^2 G(qV)}{d^2V},
\end{equation}
{where $G(qV)=G(qV;\omega=0)$ denotes the DC differential conductance.} {Such a method offers} an interesting and direct access to $q$, the r.h.s.\  can be determined from $G(qV)$. {Contrary} to the previous one, it does not necessitate a ZBA. Even more, it is valid in a more accessible regime where heating can be {reduced}: $qV,\omega\ll T$. In this case, one needs to access weak non-linearities, the second derivative of the DC conductance on the r.h.s.\ of Eq.\ (\ref{expansion_G}) being now taken at $V=0$. This is experimentally very promising and advantageous compared to the shot noise result, Eq.\ (\ref{noise}), which requires at least $qV\gg  T$ \cite{note_condition}.

A third regime corresponds to low voltage with respect to frequency, $qV\ll \omega$, without specifying $T$. This is a linear regime, where the AC conductance depends only on $\omega$. In particular, if $I$ is odd with respect to voltage, Eq.\ (\ref{reG}) reduces simply to
\begin{equation}
\label{high_frequency}{\rm Re}\,G(0;\omega)\simeq \frac q{\omega}I(qV=\omega).
\end{equation}
 In order to get a non-trivial dependence on frequency, $\omega/q$ has to be inside the domain of nonlinearity of $I$. {Then $q$ can be found as a scaling factor by comparing the linear AC conductance ${\rm Re}\,G(0;\omega)$
with the nonlinear DC current plotted versus $V=\omega/q$ and divided by $\omega/q$.}

{\bf Photo-assisted Current (PAC):}
Let us now consider the case of a time-dependent ``effective'' voltage $\tilde{V}(t)$. It may arise either from a time-dependent gate voltage, barrier height, voltage bias, or from all together. We focus, for simplicity, on a periodic variation with a unique period {$2\pi/\Omega$}, thus $f(t)=\sum f_n e^{i\omega_n t}$ where $f(t)$ stands for $|\gamma(t)|$, $\tilde{V}(t)$ or $E(t)$, and $\omega_n=n\Omega$. Multiple periods or more general time-dependence will be treated elsewhere \cite{ines_eugene}. Before expressing the PAC, let us specify $E_n$ in some typical situations.
First, letting $|\gamma(t)|=1$ and $\tilde{V}(t)=2V_{\Omega}\cos(\Omega  t)$,
thus ${\tilde V}_n=V_{\Omega}\delta_{n-1}$ for $n>0$,
\begin{equation}\label{bessel}E_n=J_{n}\left(\frac{2qV_{\Omega}}{\Omega}\right),
\end{equation} where $J_n$ are Bessel functions.
Second, for $|\gamma(t)|=1$ and sharp pulses $\tilde{V}(t)=\phi_0\sum_n\delta(t-2\pi n/\Omega)$, we obtain:
{$E_n=\sin(q\phi_0/2)/(\pi n-q\phi_0/2)$}.

We could also consider a situation where the tunneling amplitude is periodic in time while the applied voltage is constant. It is possible, {as in edge states for instance,} to modulate the gate voltage: $V_g(t)=V_g+v_g\cos \Omega t$. For weak $v_g$, {one can expand}: $\gamma (t)=1+v_g[z  e^{i \Omega t}+ z' e^{-i \Omega t}]$ where $z$ and $z'$ are complex coefficients, {thus}
\begin{equation} E_n= \delta_n+v_g(z \delta_{n-1}+z' \delta_{n+1}).\label{Engate}\end{equation}

Let us now compute the time-dependent average current, $I(t;qV)$, whose dependence on all $E_n$ is not specified to avoid confusion. Here we focus on its DC component:
{$I_0(qV)=(\Omega/2\pi)\int_0 ^{2\pi/\Omega} I(qV;t) dt$,}
the easiest to measure experimentally. It can be written as:
\begin{equation}
\label{I0}
I_0(qV)= \sum_n |E_n|^2 I(qV- \omega_n).
\end{equation}
Thus the PAC is a superposition of the DC currents at translated voltages $V-n\Omega/q$, due to the absorption
of $n$ photons, with a probability for each process given by $|E_n|^2$.
This generalizes a main result of the PAT \cite{tunnel,tucker_two} obtained for {the special case:
for $q=e$}, a tunneling amplitude constant in time, and an AC voltage, so that $E_n$ are given by
the Bessel functions, Eq.\ (\ref{bessel}).
We propose two alternatives to use Eq.(\ref{I0}) for the determination of $q$.
 The first one is suited for the case of a ZBA, taking:
\begin{eqnarray}\label{second_I}
\partial^2_ { V}I_0(qV)&=&\sum_n |E_n|^2 \partial_VG(qV-\omega_n).
\end{eqnarray}
If $\partial_VG$ is peaked around zero, this expression has peaks spaced by $\Delta V=\Omega/q$. Here $T$ needs to be small enough compared to $\Omega/q$, such that it does not smear them.

 The second alternative does not require any ZBA, neither low temperatures, thus is less restrictive. It is obtained by taking the differential with respect to the AC components of $\tilde{V}$, using Eq.\ (\ref{properties}):
\begin{equation}\label{secondI}
\frac{\omega_n^2}{q^2}\frac {\delta^2 I_0(qV)}{\delta {\tilde V}_n\delta {\tilde V}_{-n}}=-2I_0(qV)+I_0(qV+\omega_n)+I_0(qV-\omega_n).
\end{equation}
{Here we have used a general exact property of $E_n$ (assuming $|\widetilde{\gamma}(t)|$ does not depend on $\tilde{V}$):
 \begin{eqnarray}\label{properties}
\frac {\delta E_n}{\delta \tilde{V}_m}&=&-\frac{q}{\omega_m}{E}_{n-m}.
\end{eqnarray}}{Eq.(\ref{secondI})} can be also used to generate higher-order derivatives. It is powerful as it is valid still at finite and arbitrary $\tilde{V}$. It has not been derived so far within the PAT studies, even for noninteracting electrons. We see that once $I_0$ is measured as a function of $V$ for a given $\omega_n$, comparison of both sides allow to determine $q$.

Nevertheless this method can be simplified further as follows. One starts from a DC voltage $V$ to which a small ac modulation is added, $\tilde{V}(t)=v_{\Omega}e^{i\Omega t}+v_{-\Omega}e^{-i\Omega t}$. Then one measures the PAC $I_0(qV)$, and takes its second differential with respect to
$v_{\Omega}$ and $v_{-\Omega}$. According to Eq.\ (\ref{secondI}), in the limit of vanishing $\tilde{V}$ the currents appearing on the r.h.s.\ are nothing but DC currents:
\begin{equation}\label{secondIbis}
\frac{\Omega^2}{q^2}\left.\frac {\delta^2 I_0(qV)}{\delta { v}_{\Omega}\delta { v}_{-\Omega}}\right|_{\tilde{V}\rightarrow 0}=-2I(qV)+I(qV+\Omega)+I(qV-\Omega),
\end{equation}
A similar formula was derived for $q=e$ { in Ref.\cite{tucker_two}}. {The r.h.s.\ can be obtained from the DC current by
shifting the argument $V$ by $\pm\Omega/q$. Thus, for a given frequency $\Omega$ the quasi-particle
charge $q$ is a scaling parameter which makes both functions of $V$ to coincide.} {For this, at least of the three voltages, $V$, $V+\Omega/q$, and $V-\Omega/q$, has to be in the domain where the DC current is nonlinear. }

{\bf Application to the FQHE and quantum wires} Two edge states in the FQHE are described by a total Hamiltonian $\mathcal{H}_0$ in Eq.\ (\ref{Hamiltonian}), while tunneling between them can be induced by pinching off a quantum point contact (QPC). We do not require $\mathcal{H}_0$ to be decoupled into two Hamiltonians, but only to conserve {the charges of both edges,
$Q_1$ and $Q_2$}. This is a strength of our approach: even Coulomb interactions between the edges, usually present at least within the QPC, can be incorporated in $\mathcal{H}_0$. Another strength is that the knowledge of $\mathcal{H}_0$ is not required in order to test the charge in the bulk, or the effective one, contrary to methods using the explicit expressions for the current or the noise. {Moreover, the charge $q$ which appears in Eq.(\ref{tunneling-t0}) is robust against the coupling to an ohmic environment, contrary to the DC conductance without tunneling or the exponents controlling the power law behavior (of the current and its fluctuations)\cite{note_env}.}

We have shown that different regimes can be investigated, depending on energy scales and technical limitations. An additional strength, in comparison to the measurement of the noise in Eq.\ (\ref{noise}) which requires $qV\gg T$ \cite{note_condition}, is that we have proposed three methods in the regime $qV\ll T$, {which reduces} heating. {Quite often, the nonlinearity domain of $I(qV)$ corresponds rather to $qV\gg T$.}

Let us focus on these proposals.
 Among them, two are based on the AC conductance; either it is measured at low frequencies compared to $T$, where one needs to access weak non-linearities of the DC conductance, see Eq.\ (\ref{expansion_G}), or, if frequencies higher than $qV$ can be reached, one can use directly Eq.\ (\ref{high_frequency}). The third one,
based on the PAC, can exploit Eq.\ (\ref{secondIbis}), which requires here that $V+|\Omega|/q\gg T$. If one needs in addition $T\gg qV$ to reduce heating, one must fulfill $|\Omega|\gg T$. {This is easily accessible as $\Omega$ can be varied from the RF up to the optical spectrum, thus can be made easily larger then many kelvins. Notice that in this regime, the r.h.s.\ of Eq.\ (\ref{secondIbis}) simplifies to $-2qVG+I(\Omega)+I(-\Omega)$.} {Besides being a zero-frequency measurement}, this method has another advantage compared to the AC conductance: one can control the time-dependence of the tunneling amplitude, through the gate voltage, while keeping only a DC applied voltage, yielding for instance the amplitudes $E_n$ in Eq.\ (\ref{Engate}). This has less drawbacks compared to an AC voltage which needs to propagate along the edges.

We have assumed that only one process with a given charge {$q$} dominates at once. It turns out that {the value of $q$} depends on the regime one considers. For simple fractions, $\nu=1/(2m+1)$, the tunneling charge is $q=\nu e$
(or $q=e$) at large (respectively, low) energies compared to a non-universal energy scale $T_B$.
 One can view this as a bunching of $2m+1$ quasi-particles below $T_B$. For other series of filling factors, the question is still controversial. In particular, Heiblum's group claimed \cite{heiblum_bunch,heiblum_two_third} that bunching of the fundamental quasi-particles occurs at low energies \cite{sassetti_10}.
 For instance, at $\nu=5/2$ (or $\nu=2/3$) one measures $q=e/4$ (respectively, $q=e/3$) at high energy, while one observes rather $q=e/2$ (respectively, $q=2e/3$) at low energies. Such bunching was though excluded by Radu {\it et al} \cite{radu_exp}. Indeed, Heiblum's group {used the expression for noise of} non-interacting particles with a charge $q$, which has to be taken with caution. Our proposals, which do not require any explicit expressions for the current neither noise, would be extremely useful to solve these controversies, { and to show with a clear evidence the shift between different charges.}

{At resonance, i.e., when the gate voltage is adjusted to suppress the tunneling amplitude for one quasi-particle, it was predicted that for $\nu=1/(2n+1)$ tunneling of two quasi-particles should dominate,}
thus doubling $q\rightarrow 2q$ \cite{kane_fisher_noise}. This was not observed in experiments, $q=\nu e$ being measured \cite{charge_hall},
which was explained in Ref.\ \cite{ines_resonance}: still one-quasi-particle tunneling dominates.

Another setup concerns junctions between edges at different filling factors $\nu_1$ and $\nu_2$, for which it was predicted \cite{ines_schulz,ines_ann}, for simple fractions, that the junction behaves as an edge with effective ``filling'' $\nu=2\nu_1\nu_2/(\nu_1+\nu_2)$. If one couples a Fermi liquid to an edge state \cite{chang}, one sets formally $\nu_1=1$. This has motivated other theoretical \cite{halperin_sandler} and experimental investigations \cite{roddaro_09}. The DC conductance, expected to be given by $e^2\nu/h$, is nevertheless difficult to observe as it is sensitive to the geometry, depletion, backscattering, etc. One rather expects the effective charge $\nu e$ to be more robust, thus our methods to be of great relevance. Yet another setup consists in coupling an edge state to a superconductor \cite{theo_supra_hall,exp_supra_hall}, where the effective charge is given by $q=2\nu e$ for simple fractions \cite{ines_ann}.


Our results apply to quantum wires (QWs) or carbon nanotubes (CNTs). It was predicted that an injected electron splits into two solitonic charges \cite{ines_ann,pham,frac}. For tunneling junctions as well as weak barriers, the charge $q$ appearing in Eq.\ (\ref{noise}) was expected to be respectively given by $e$ and $Ke$ \cite{kane_fisher_noise}, where $K$ is a parameter depending on interactions, and is roughly the analog of $\nu$. However, contrary to the FQHE, the coupling to the reservoirs \cite{ines_schulz,ines_ann,maslov_g} changes $q=Ke$ to $e$ in Eq.\ (\ref{noise}) \cite{ponomarenko_features}, unless high-frequency noise is considered \cite{trauzettel_04}. Even though measuring $q=e$ looks trivial, it would be an interesting check of the underlying model, adopted so far by the community.
More generally, provided that $\mathcal{H}_0$ is quadratic in the bosonic fields, describing arbitrary range, form and inhomogeneities of the interactions and parameters, we can show that the weak backscattering term in Eq.\ (\ref{tunneling-t0}) contains $q=G_0,$ the dimensionless DC conductance {of a system described by} $\mathcal{H}_0$ only \cite{note_DC}. This could be relevant for edge states with {a quadratic action}, for instance if the filling factor in the QPC is different from that in the outer edges.

There is a limitation on $\omega$ which matters only for the methods based on the AC conductance and for 1-D systems. The current operator in the electrodes is, strictly speaking, different from the tunneling current operator in Eq.(\ref{IT}). When its average is measured at some distance $L$, the AC average values of both operators coincide only at low frequencies, $\omega \ll \omega_L=v/L$ where $v$ is the plasmon velocity, otherwise the propagation from the tunneling junction to the contacts have to be incorporated \cite{bena_07,safi_ac}. This condition is easily fulfilled for typical QWs or CNTs whose length is {of the order of $10\mu m$ or less, thus $\pi\omega_L\simeq 100 GHz$--$1 THz$}
depending on $v$, while $\omega$ cannot exceed $20 GHz$. {It is as well the case in the FQHE, where for $L\simeq 5 \mu m$, and $v\simeq 5. 10^5 m/s$, $\omega_L\simeq 100 GHz$.} One does not expect the limitation on $\omega$ to hold in higher dimensions, {where
the plasmon spectrum typically has a large gap}.
Notice also that QWs and CNTs are examples where the non-linearity domain of $I$ is not controlled only by $T$. At $V\ll \omega_L/e$, even if $T\ll eV$, the noninteracting leads impose their ``linearity'' \cite{ines_ann,safi_ac}.

{{\bf Conclusion:}}
We have considered a non-linear tunneling junction between two arbitrary systems in any dimension, with any interactions or disorder, excluding only a super-current. We have proposed many methods to access the charge $q$ of carriers. The main advantages of our proposals is that the underlying model does not have to be specified, that they are based on a current measurement, and that many of them are applicable at low frequency or/and voltage. They are based on two quantities, each of them can be exploited into two different ways: the AC conductance \cite{AC}, and the PAC when either the voltage or the tunneling barriers, or even both, have a periodic time-dependence.

On one hand, the methods based on the AC conductance are proposed in three situations. The first requires a ZBA, $\omega$ of the order of $qV$, and at least one of $V\pm \omega/q$ in the nonlinearity domain of $I$. Singularities of Eq.\ (\ref{reG}) at $\omega=\pm qV$ gives access to the charge. This feature has been observed in the FF noise for simple fractions in the FQHE \cite{chamon_noise,bena_07,crepieux_photo}. Here we have established its universal existence in the dissipative AC conductance, which yields the asymmetric part of the FF noise. This is related also to the universal appearance of a similar feature in Eq.\ (\ref{I0}).

 The second regime does not require any ZBA, and holds even at $\omega,qV\ll T$, the easiest regime to attain experimentally, which reduces heating effects too. The weak non-linearities of $G(qV)$ in Eq.\ (\ref{expansion_G}) give access to $q$.

 The third regime is linear with respect to $V$, as it corresponds to $\omega\gg qV$, where $G(qV;\omega)\simeq G(0;\omega)$. The charge $q$ can be deduced easily from Eq.\ (\ref{high_frequency}). One needs to vary $\omega/q$ in the domain where the DC current is nonlinear, which depends on the system at hands. In the FQHE, it corresponds to $\omega/q\gg T$, accessible experimentally as well. Typically, a voltage of $40 \mu eV$ corresponds to $\simeq 10 GHz$, and a temperature of $50 mK$ to $\simeq 1 GHz$. Technically, it has become possible to reach $\omega\simeq 20 Ghz$, thus greater than both the typical $V$ and $T$.

  On the other hand, let us recall the methods based on the PAC, expressed through the general formula (\ref{I0}). Here $\Omega$ corresponds to the period {$2\pi/\Omega$ of the applied voltage, and} can be made much greater than $\omega$ above, at which the current is measured.
The first method requires a ZBA and $T\ll qV$ such that the spikes in Eq.\ (\ref{second_I}) are not smeared out. As noticed before, this is related to the similar feature in the AC conductance, Eq.\ (\ref{reG}), and in the FF noise \cite{ines_eugene}. Nevertheless, $I_0$ is much easier to measure compared to the FF noise, being in addition at zero frequency. The second method based on the PAC, using Eq.\ (\ref{secondIbis}), does not necessitate a ZBA, but an arbitrary non-linear behavior. Even more, it does not require low temperature compared to both $\Omega$ and $V$, but that only one of $V,V\pm \Omega/q$ to be in the nonlinearity domain of $I$.

Non-linearity is indeed a crucial ingredient in all our methods, {which are particularly promising for the FQHE.}
They could be useful for super-conducting structures as well; the noise in Eq.\ (\ref{noise}) being very weak, more complicated setups were adopted to measure $q=2e$ \cite{sanquer_noise}. Alternatives for linear systems will be proposed elsewhere \cite{ines_eugene}. We will also express both the current average and its fluctuations for any time dependence of the voltage and the tunneling amplitude in linear or nonlinear systems.

Acknowledgments: I. S. is grateful to R. Deblock, B. Dou\c cot, D. Est\`eve, M. Goerbig, P. Joyez, F. Pierre and P. Simon for fruitful discussions and encouragements.

\begin{thebibliography}{99}
\bibitem{note_condition}
Indeed, the condition is even stronger for the edge states,
where the voltage needs to be higher. See Ref.\ \cite{trauzettel_04}.
\bibitem{heisenberg}
Operators between brackets are implicitly taken in the Heisenberg representation.
\bibitem{AC}
 \bibinfo{note}  Let us recall its meaning \cite{safi_ac}. Suppose one adds to $V$ a perturbation:
$\delta \tilde{V}(t)=v_{\omega}e^{-i\omega t}+{\rm h.c.}$. For
$v_{\omega}$ weak enough, i.e.,  if
$q\delta \tilde{V}/\omega\ll 1$ for $\omega\gg qV$, and $\delta\tilde V\ll V$
otherwise, one has
$I(qV;t)=I(qV)+[G(qV;\omega)v_{\omega}e^{-i\omega t}+{\rm h.c.}]$. Notice that by measuring $G(qV;\omega)-G(0;\omega)$\cite{safi_ac} one avoids undesirable effects independent on $V$.
\bibitem{note_env}
For simple fractions $\nu$, it was shown that still $q=\nu e$, while the exponent is controlled by $\nu/(1+r\nu)$, where $r$ is the dimensionless resistance of the environment. See I. Safi and H. Saleur, Phys. Rev. Lett. {\bf 93}. (2004), 126602.
\bibitem{frac}
Calling this phenomena ``a charge fractionalisation'' has to be used with caution.
\end{thebibliography}

\end{document}